# Conception of Quantum Cosmology

© I.N. Taganov[1,2]

1. Russian Geographical Society, Saint Petersburg, Russia
2. taganov.igor@mail.ru

**Abstract.** Quantum cosmology describes universe as a relativistic object with an evolution defined by an equation for the energy density corresponding to the least action principle: $\rho_E = const/\tau$ (Taganov, 2008). In quantum cosmology this equation plays the same role as the Planck equation does in quantum physics. Universe evolution consists of not only the space expansion but also a deceleration of the course of physical time. Durations of all processes, measured by decelerating physical time, are always longer, than corresponding durations, measured by a scale of the invariable uniform Newtonian time. Quantum interpretation of the redshift phenomenon enables to estimate the Hubble parameter by means of fundamental physical constants: $H = 9G\hbar/16c^2 r_e^3 = 1.970 \cdot 10^{-18}$ s$^{-1}$ (61.6 km/s/Mpc). In the course of evolution, the universe retains the self-similarity defined by the constancy of the characteristic scale relations for micro- and mega worlds with an average fractal dimension of the typical cosmic large-scale structures $D = 2$. Quantum cosmology advances new relations for the microwave background parameters, apparent stellar magnitudes and redshifts; formulae for the cosmological increase of the macroscopic space- and time- characteristics and evaluation of quasar redshifts and luminosities.

## Introduction

Quantum cosmology could be said to have begun with Max Plank' proposal in the conclusion of his legendary presentation in Academy of Sciences in Berlin on May 18, 1899 to introduce the "natural units" of measurement, basing on his new quantum constant. Plank' idea, however, got no support from his contemporaries, and it was buried in oblivion for more than half a century until in the 1950s John Wheeler rediscovered Planck' fundamental length in his "geometro-dynamics". In 1958 Nikolai Kozyrev achieved an important heuristic result introducing first global cosmological quantum parameter - the "course of time constant" $e^2/h$ [16], but like Planck he had not many followers. Despite occasional criticism, cosmology continued to use Newton-Einstein gravitation theory, abandoning for a long time an idea of the search for specific relativistic and quantum laws of mega-world. This was by no means because the failure to realize limited prospects of a mega-world theory based on Newton-Einstein gravitational equations and thermodynamics. The quest for specific quantum mega-world laws was inhibited, until the last quarter of the 20th century, by inferior, compared to quantum physics, amount of reliable quantitative data from observations of distant cosmic structures. An important stimulus for progress in quantum cosmology was the discovery of fractal geometry of the universe large-scale structures. It appeared that fractal dimension of the typical universe large-scale structures $D = 2$ is the same as the dimension of a fractal micro-particle trajectory described by quantum mechanics.

A conception of relativistic quantum cosmology can be formed from the following basic ideas [13, 14]:
1. Quantum cosmology, as a part of quantum physics, is specifically concerned with describing discrete rather than continuous space distributions of matter. Unlike classical cosmology, quantum cosmology should therefore use mass and energy densities in the form of extensive characteristics of finite volumes and masses as opposed to intensive densities represented by continuous physical fields. Intensive parameters of the mass or energy density type are basically inadequate for fractal galaxy distributions in the large-scale universe structure, being explicitly dependent on the averaging volume and tending to zero as the volume increases in the galaxy distributions with high lacunarity.
2. Like the quantum micro-world theory, quantum cosmology can rely on the specific mega-world equation having a methodological unity with quantum postulates of Planck and de Broglie. Early attainments of quantum physics were associated with the use of quantum postulates long before the formalisms of wave function and matrix mechanics were developed. For quantum cosmology is especially important the following unique methodological advantages of the Planck equation:
- a laconic form of the least action principle
- coordination and interrelation of the motion characteristics that provides a union of discrete corpuscular and continuous wave descriptions of the micro-world processes.
3. Quantum cosmology has to be relativistic theory firstly since the universal constancy of the speed of light is a major result of quantum photon physics. Secondly, because the constancy of the speed of light is the basic principle of contemporary length and time standards. On the other hand, relativistic nature of quantum cosmology is not governed by Lorentz transformation group or consequences of the principle of the speed of light constancy in various inertial reference systems with related relativistic mechanics. Of particular importance for cosmology are the following features of relativistic ideology:
- universal constancy of the speed of light everywhere in space and in all epochs of the universe evolution
- quantum cosmology requires a relativistic interpretation of the mass and energy relations in processes where the energy is comparable with energy equivalents of the rest masses.

Let us consider the basic laws, which define main equations and dimensional analysis in classic and quantum physics:

**MACRO WORLD:**   Newton law: $\varphi_g = \Phi_g/m = -G/r$   Coulomb law: $\varphi_e = \pm e/r$
**MICRO WORLD:**   Planck law: $\varepsilon = \hbar/\Delta t$   de Broglie law: $p = \hbar/\lambda$

One may see that all these basic laws have the universal form of a power-law relation of the first negative order. Quantum cosmology demonstrates that for mega world, i.e. for cosmological distances much bigger than the typical galaxy sizes and corresponding time intervals, the analogous power-law relation for average energy density is valid:

**MEGA WORLD:**   $\rho_E = k_T/\tau$   $k_T = 2c^2 H/3\pi G = 3\hbar/8\pi r_e^3 = 5.630 \cdot 10^2$ J s cm$^{-3}$

Rapid progress of astronomical instruments and computer technologies in the last quarter of the bygone 20[th] century enriched cosmology with numerous observational data on the structure of distant cosmos. In spite of it, the amount of quantitative parameters of the entire integral universe grows slowly. Now we may fruitfully discuss only four Key Cosmological Parameters, which can be estimated by different independent observational methods (the Table 1):
- Hubble parameter and average universe mass density (from 1930s)
- Energy density and temperature of CMB - cosmic microwave background radiation (from 1970s)
- Fractal dimensions of the universe large-scale structures (from 1980s)

*Table1. Key Cosmological Parameters*

| Key Cosmological Parameters | Observations | Quantum Cosmology Estimations (Taganov, 2008) |
|---|---|---|
| Hubble parameter (km/s/Mpc) | (Riess et al, 2004) $H = 65 \pm 7$ (Sandage et al, 2006) $H = 62.3 \pm 1.3$ | $H = 9G\hbar/16c^2 r_e^3 = 1.970 \cdot 10^{-18}$ s$^{-1}$ = 61.6 km/s/Mpc |
| Average mass density | $\rho_m = (5 \div 10) \cdot 10^{-30}$ g cm$^{-3}$ | $\rho_m = 4H^2/9\pi G = 9\hbar^2 G/64\pi c^4 r_e^6 = 8.217 \cdot 10^{-30}$ g cm$^{-3}$ |
| CMB energy density and temperature | $\rho_{CMB} = 4.19 \cdot 10^{-13}$ erg cm$^{-3}$ $T_{CMB} = 2.728 \pm 0.004$ K | $\rho_{CMB} = \rho_m e^4/\hbar^2 = 3.929 \cdot 10^{-13}$ erg cm$^{-3}$ $T_{CMB} = (\rho_{CMB}/\sigma)^{1/4} = 2.684$ K |
| Fractal dimension of the universe large-scale structures | $\rho_m \propto r^{-1}$ $D = 2 \pm 0.2$ | $\rho_m = (3\hbar/8\pi c r_e^3) r^{-1} = 1.878 \cdot 10^{-1} \cdot r^{-1}$ g cm$^{-3}$ $D = 2$ |

Gravitational constant $G = 6.673 \cdot 10^{-8}$ cm$^3$ g$^{-1}$ s$^{-2}$; Planck constant $\hbar = h/2\pi = 1.055 \cdot 10^{-27}$ erg s; speed of light in a vacuum $c = 2.998 \cdot 10^{10}$ cm s$^{-1}$; charge of electron $e$ ($e^2 = 2.307 \cdot 10^{-19}$ g cm$^3$ s$^{-2}$); classic electron radius $r_e = e^2/m_e c^2 = 2.818 \cdot 10^{-13}$ cm; Stefan-Boltzmann constant $\sigma = 7.566 \cdot 10^{-15}$ erg cm$^{-3}$ K$^{-4}$.

Besides numerous successful qualitative predictions and elegant mathematical analyses, cosmology based on Einstein-Friedmann equations exposes strange Paradox of Theoretical Uncertainty: the absence of theoretical estimations of the Key Cosmological Parameters even with symbolic accuracy. Many alternative models existing in contemporary cosmology reveal the same paradox. Quantum cosmology gives an example of successful defeat of the Paradox of Theoretical Uncertainty (Table 1).

# 1. Quantum equation of the universe evolution

Cosmology describes expanding universe with non-stationary metrics defining intervals of the type:

$$ds^2 = -c^2 d\tau^2 + a^2 dr^2 \qquad (1)$$

The dimensionless scale-factor $a(\tau)$ defines variations of space intervals relative to the constant standard $R_0$: $r(\tau) = a(\tau) R_0$. In the space-time with interval (1), the speed of light defined as coordinate velocity at the geodesic line with zero interval ($ds = 0$) appears as variable:

$$dr/d\tau = c/a(\tau) \neq const \qquad (2)$$

The speed of light is invariable and can be regarded as a universal constant only at the world-lines with Minkowski metric using Newtonian time $t$ and determining the interval:

$$ds^2 = -c^2 dt^2 + dr^2 \qquad (3)$$

Therefore, the use of non-stationary metrics with interval (1) in cosmological models contradicts a conception of the quantum photon physics asserting universal constancy of the speed of light in vacuum. The photon velocity equation (2) can be transformed using the scale-factor $a(\tau) = r(\tau)/R_0$ to: $dr/d\tau = c R_0 / r$. A condition of the constancy of the speed of light on the geodesic line then looks like: $r\, dr/d\tau = c R_0 = const$. Integrating this equation with initial condition $\tau = 0 : r = 0$, we obtain the relation: $r^2 = 2 R_0 c \tau$. Thus, a condition of the constancy of the speed of light in a non-stationary universe with arbitrary time-dependence of the scale-factor results in the following relation for space- and time intervals:

$$r^2 \propto \tau \qquad (4)$$

For density dependence of the chemical potential: $\mu \propto \rho^n$ the pressure-density relation is defined by the equation of state (see e.g. [8, 9]):

$$p \propto \rho^{n+1} \qquad (5)$$

For a relativistic matter $n = 1/3$ and the total energy of the matter with the equation of state (5) is defined by the relation: $E = -\dfrac{3n-1}{5n-1}\dfrac{Gm^2}{r} = 0$ with potential gravitational energy: $U_G = -\dfrac{3n}{5n-1}\dfrac{Gm^2}{r} = -\dfrac{3Gm^2}{2r}$. Total energy of the relativistic matter, including the energy equivalent of the rest mass, looks like:

$$E = mc^2 + E_u - 3Gm^2/2r = 0 \qquad (6)$$

Here $E_u$ is the internal energy including kinetic energy of moving subsystems. As virial theorem affirms that: $E_u = -U_G/2 = 3Gm^2/4r$, the relation (6) can be transformed to:

$$mc^2 - 3Gm^2/4r = 0 \qquad (7)$$

From this energy balance one can derive the mass density $\rho_m = m/V = 3m/4\pi r^3$ defined as an extensive parameter for finite volume and mass: $\rho_m = c^2/\pi G r^2$. Inserting Eq. 4 into this relation one can get:

$$\rho_m \tau = const \qquad (8)$$

In relativistic methodology, the energy conservation law generalizes the mass conservation law, accounting for example possible mass and energy transformations with changing mass defect in the structures of interacting elements. Conversion of the mass density to energy density $\rho_E$ with the use of the mass energy equivalent: $\rho_E = \rho_m c^2$ will bring Eq. 8 into the form:

$$\rho_E \tau = const \tag{9}$$

It must be remembered that densities in Eqs. 8, 9 are defined, as opposed to intensive parameters representing continuous physical fields, as the average densities of a finite structure, i.e. as ratios of extensive characteristics: energy, mass and volume. The use in Eqs. 8, 9 mass- and energy densities defined as extensive average characteristics of finite structures allows their application in description of the evolution of heterogeneous matter- and energy distributions in the universe. However, the extensive nature of average mass- and energy densities prevents their use in the differential thermodynamic relations.

## 2. Quantum cosmological model

In cosmology the term "cosmological model" is used to name equations describing the scale-factor variation in the process of universe evolution. A doctrine of "expanding" space of the universe $\Delta l(\tau) = a(\tau) \Delta l_0$ with monotone increasing scale-factor $a(\tau)$ together with a condition of the constancy of the speed of light: $\Delta l = c \Delta \tau; \Delta l_0 = c \Delta \tau_0; c = const$ lead to the relation: $\Delta \tau(\tau) = a(\tau) \Delta \tau_0$. This relation suggests that time also "expands" along with the space in the course of universe evolution. While a term "space expansion" is a common cosmological term today, the somewhat clumsy term "time expansion" is better to replace with a more accurate term "deceleration of the course of time". The course of time is defined as a value $\Delta \tau^{-1}$, converse to the chosen time standard $\Delta \tau$. Increasing time standard corresponds to decreasing course of time and thus to deceleration of the course of time. The course of time concept was probably first formulated by Einstein and Minkowski in their pioneering works in the relativity theory. The term "course of time" was later favored by J. Synge [12] and N.A. Kozyrev [15].

The decelerating, expanding time cannot be Newtonian time commonly used by natural science as invariable homogeneous continuum. The time displaying a deceleration in the process of universe evolution is referred to in this article as "physical". The term "physical time" is justified by analogy with the term "physical vacuum" used by quantum physics instead of the old classical concept of "emptiness" as an abstract three-dimensional mathematical continuum. Quantum physics defines vacuum state by fluctuations of interacting quantum fields. These fluctuations correspond to zero-oscillations in quantum mechanics and govern multiple transformations of virtual micro-particles resulting, in particular, in physical vacuum polarization. The fluctuation spectrum change and vacuum polarization in volumes with electro-conducting boundaries are made evident by Casimir macroscopic forces, independent of masses, charges or any other coupling factors.

Since modern physics accepts a conception of non-stationary space-time, a principle of the constancy of the speed of light and quantum postulates, the decelerating time rightfully can be referred to as "physical". Similar to the physical vacuum theory, our conception of the cosmological deceleration of the course of time is substantiated with relativistic and quantum ideology. Physical time is henceforth symbolized by $\tau$, Newtonian time by $t$, with $a' = da/d\tau$ and $\dot{a} = da/dt$.

A condition of the constancy of the speed of light allows to obtain from equivalence of (1) and (3) the coupling equation for $\tau$, and $t$:

$$d\tau/dt = a \tag{10}$$

Notice that this equation can be derived with the condition $dr = 0$, i.e. for *unmoving objects*. From the Eq. 10 it follows that all characteristic intervals of physical time $\tau$, used in non-stationary metrics with interval (1) and monotone increasing scale-factor, will grow with respect to the uniform and invariable Newtonian time scale. The equation describing the scale-factor growth can be derived after transformation of (4) using the scale-factor definition $r = aR_0 : (aR_0)^2 \propto \tau$. Differentiation of this relation with respect to $\tau$ leads to: $a'a = const$ and differentiation of this equation in its turn gives:

$$a''a + a'^2 = 0 \qquad (11)$$

This equation of quantum cosmological model can be also derived from the Eq. 9. Taking into account the change of dimension scales of basic units in a non-stationary universe: $[l] = a[l_0]$, $[\tau] = a[\tau_0]$, $[m] = a[m_0]$, Eq. 9 can be written as: $\rho_E \tau = \rho_m c^2 \tau = m_0 \tau / a^2 l_0^3 = const$, or $a^2 = const \cdot \tau$. Differentiation of this relation with respect to $\tau$ leads to: $a'a = const$ and repeated differentiation gives the Eq. 11.

Equation (11) describing scale-factor evolution in quantum cosmological model can be represented in traditional for theoretical cosmology form, defining cosmological deceleration parameter $q_\tau$ for physical time:

$$q_\tau = -aa''/a'^2 = 1 \qquad (12)$$

In addition to Eqs. 11, 12 the quantum cosmological model should include the proper frame of reference and initial conditions. Instruments for observations whose functions are described by either quantum or classical physics play an important role in cosmology. In quantum physics a conception of "observational relativity" is used to underline the leading role of "classical instrument" in quantum theory [4]. In cosmology a function of "classical instrument" of quantum physics executes the frame of reference with special emphasis on zero-time reference point. As quantum physics methodology depends on the properties of "classical instrument", quantum cosmology relies on a frame of reference.

Processing their observational data for most bright stars in the galaxies during the 1920s, Knut Lundmark and Edwin Hubble calculated the spectral shifts from the same formula as used by astronomers today: $z = \Delta\lambda = (\lambda_p - \lambda^*)/\lambda^*$. Here $\lambda^*$ is a standard laboratory wavelength corresponding to observed spectral line $\lambda_p$. The use of the scale-factor allows writing this formula as:

$$z = \Delta\lambda = \frac{\lambda_p - \lambda^*}{\lambda^*} = \frac{\lambda^* a(t_p) - \lambda^* a(t_r)}{\lambda^* a(t_r)} = \frac{a(t_p)}{a(t_r)} - 1 \qquad (13)$$

Here $t_r$ is the moment of radiation emission. The index "p" hereinafter identifies present-day values of cosmological parameters. For expanding universe with $t_p \geq t_r$ the relation $a(t_p) \geq a(t_r)$ holds and spectral shift is "red", i.e. spectral lines shifting towards the long-wave side of the spectrum. Since the days of Lundmark and Hubble, cosmological redshifts have been computed using observed spectral wavelength $\lambda_p$ at the reception time i.e. at our epoch. In comparing redshifts in spectra of various cosmic objects, the value $\lambda_p = \lambda^* a(t_p)$ is a variable with $\lambda^* a(t_r) = const$ determined by the laboratory standard. As the scale-factor allows the arbitrary multiplier, it is possible to set $a(t_r) = 1$. In this case, the expression (13) becomes:

$$\Delta\lambda + 1 = z + 1 = a(t_p) \qquad a(t) = z + 1 \qquad (14)$$

The emission reception time $t_p$ in this relation is used as the current time and independent variable. The emission time: $t_r = 0$ is the zero-time reference point.

It would be convenient to have a single, "absolute" time scale with zero-time reference point at a hypothetical initial moment of the universe evolution. However, we have no reasons to find emission times for various space objects on this time scale. One can only assume that: $a(t_p) = 1$, believing that zero-time reference point corresponds to: $a(t_p = 0) = 0$. For this frame of reference Eq. 13 defining the redshift looks like:

$$Z(t) = \frac{\lambda_p^* - \lambda_p^* a(t_r)}{\lambda_p^* a(t_r)} = 1/a(t_r) - 1 \qquad a = 1/1+Z \qquad (15)$$

This formula uses emission time $t_r$ as a current time and independent variable. Initial countdown for $Z(t)$ corresponds to the hypothetical universe evolution start-time $t=0$. While often proving useful in transformations of cosmological model equations, the expression (15) is inconsistent with the Eqs. 13, 14 applied to interpret the astrophysical observations of luminous cosmic objects.

These two different definitions of the cosmological redshift realize two possible scale-factor normalizations to unit: at reception time – (15) and at emission time – (14). The values Z and z may only be considered as about the same value when $Z; z \ll 1$. The formal properties of parameters Z and z are different, corresponding, in particular, to different zero-time reference points for the current time.

This analysis indicates that detailed mathematical models in quantum cosmology should use the frame of reference with zero-time reference point corresponding to the moment of emission. This frame of reference, corresponding to the redshift formula (14), also is used in the practice of astrophysical observations. The frame of reference with zero-time reference point at the hypothetical moment of the universe birth, corresponding to Eq. 15 for redshift, can be successfully used in the models, describing evolution of the whole universe as an integral object.

Solutions of Eqs. 11, 12 with initial conditions, corresponding to different renormalizations of scale-factor are the following (see e.g. [6] i. 6.125):

$$\tau = 0: a = 1, a' = a'_0 = H \qquad a = (1 + 2H\tau)^{1/2} \tag{16}$$

$$\tau = 0: a = a_0, a' = a'_0 \qquad a = (a_0^2 + 2a_0 a'_0 \tau)^{1/2} \tag{17}$$

Eq. 17 shows that a point $\{\tau = 0, a = 0\}$ is the peculiar point for Eqs. 11, 12. Using a transformation rule for derivative: $\dot{a} = da/d\tau \cdot d\tau/dt$ and Eq. 10 one can derive from Eqs. 11, 12: $\dot{a} = const$. For initial condition $t = 0: a = 1$ this equation integrates to: $a = 1 + Ht$ and after substitution of (14): $a(t) = z + 1$ it gives a common form of Hubble law with Newtonian time:

$$z = Ht \tag{18}$$

Integration of Eq. 10 with initial condition $t = 0: \tau = 0$ after substitution of the Eqs. 14, 18 in the form: $a = 1 + Ht$ gives the algebraic coupling relations for physical and Newtonian time:

$$\tau = t + Ht^2/2 \tag{19}$$
$$t = H^{-1}[(1 + 2H\tau)^{1/2} - 1] \tag{20}$$

When Eq. 20 is substituted in Eq. 18 we find the formula of Hubble law with physical time:

$$z = (1 + 2H\tau)^{1/2} - 1 \tag{21}$$

The Eq. 4 establishing quantum cosmological model (11, 12) is a laconic description of the universe evolution:

$$R^2 \propto \tau \tag{22}$$

*Square of the growing universe radius is proportional to the universe physical age.*

### 3. Cosmological scales and Hubble parameter

Estimates of the cosmological scales of time ("universe age") and length ("universe radius" or "horizon") can be deduced from the Eq. 19. Universe age is estimated in Newtonian time from the relation (14): $a = 1 + z$, assuming that $z_p = 0$ and, accordingly, $a_p = 1$. On this assumption, Eq. 18 provides Newtonian age of the universe:

$$t_p = H^{-1} = 5.081 \cdot 10^{17} \text{ s} = 16.131 \text{ Gyr} \tag{23}$$

Here and further on for the calculations of cosmological scale values a theoretical formula for the Hubble constant (29) is used. Physical age of the universe, corresponding to Eq. 23 is derived from the Eq. 19:

$$T_H = t_p + H/2 \, t_p^2 = 3/2 \, t_p = 3/2 \, H^{-1} = 7.614 \cdot 10^{17} \text{ s} = 24.1 \text{ Gyr} \tag{24}$$

The relation $R_H = cT_H$ together with Eq. 24 provides the estimate of the universe radius:

$$R_H = cT_H = 3c/2H = 2.283 \cdot 10^{28} \text{ cm} \tag{25}$$

The Eq. 7 provides the estimate of the universe mass scale:

$$M_H = 4c^2 R_H / 3G = 2c^3 / GH = 4.1 \cdot 10^{56} \text{ g} \tag{26}$$

Providing that the quantum evolution equation (8) is valid for cosmological scales (24 - 26) then: $\rho_{mp} T_H = 3 M_H T_H / 4\pi R_H^3 = 2H/3\pi G$. This relation allows to estimate the constants in Eqs. 8, 9:

$$\rho_E \tau = k_T = 2c^2 H / 3\pi G = 3\hbar / 8\pi r_e^3 = 3\hbar c^6 m_e^3 / 8\pi e^6 = 5.630 \cdot 10^2 \text{ J s cm}^{-3} \tag{27}$$

$$\rho_m \tau = \rho_E \tau / c^2 = k_T / c^2 = 6.264 \cdot 10^{-12} \text{ g s cm}^{-3} \tag{28}$$

Index "T" for quantum constant $k_T$ underlines a close connection of quantum cosmological model with a conception of physical time and the phenomenon of cosmological deceleration of the course of time.
     Physical meaning of Eq. 27 can be clarified by comparison with Planck equation, which can be interpreted as a definition of the Planck constant for a minimal action, corresponding to the electron angular momentum projection: $E\Delta t = \hbar/2 = const$ (J s), where $\Delta t = 2\pi/\omega$ is the period of oscillation associated with the micro-particle. Planck equation therefore postulates discreteness and a constancy of the *minimal action* (J s) in microcosm. The same analysis demonstrates that Eq. 27 postulates a constancy of the *action density* (J s cm-3) in mega-world. Planck equation can be transformed to a relation of the type (27) on the assumption of the existence of a finite volume $V_{Pl} > 0$ where quantum action is defined: $\rho_{EPl} \Delta t = E\Delta t / V_{Pl} = \hbar / 2V_{Pl} = const$. It may be also suggested that quantum action is defined in the same volume as the elementary charge, i.e. in the sphere with electron radius: $V_{Pl} = 4\pi r_e^3 / 3$ (here $r_e = e^2/m_e c^2 = 2.818 \cdot 10^{-13}$ cm is the classical electron radius). Equality of Planck action density and action density (27): $\hbar / 2V_{Pl} = 2c^2 H / 3\pi G$ allows one to define the Hubble constant only using fundamental constants:

$$H = \frac{9G\hbar}{16c^2 r_e^3} = \frac{9G\hbar c^4 m_e^3}{16 e^6} = 1.970 \cdot 10^{-18} \text{ s}^{-1} = 61.6 \text{ km/s/Mpc} \tag{29}$$

This theoretical value of the Hubble constant corresponds well to the Hubble parameter observational estimations. In 1927 Jorge Lemaitre using less than ten galaxy redshifts evaluated Hubble constant as 625 km/s/Mpc. Edwin Hubble himself estimated this parameter in the 1930s as 559 km/s/Mpc. In the 1940s, astrophysicists preferred the value around 200 km/s/Mpc. In 1970 - 1990s summarizing of all published data on galaxy redshifts had led to the estimation: 50 – 80 km/s/Mpc. In 2000 the multiple data of the Hubble Key Program (HKP) of Cepheid survey for galaxies at distances below 20 Mpc ($z < 0,1$) estimated the Hubble constant as $72 \pm 8$ km/s/Mpc. A recent international survey for the type Ia supernovae with redshifts $z = 0.1 \div 1$ estimated the present value of Hubble parameter as $65 \pm 7$ km/s/Mpc. In 2006 Alan Sandage, recognized redshift expert estimated the Hubble constant as $62.3 \pm 1.3$ km/s/Mpc [10].

The average mass and energy densities of the universe can be defined using cosmological scales (25, 26) and Eq. 29 for the Hubble constant:

$$\rho_m = M_H/V_H = 4H^2/9\pi G = 9\hbar^2 G/64\pi c^4 r_e^6 = 8.217 \cdot 10^{-30} \text{ g cm}^{-3} \qquad (30)$$

$$\rho_E = c^2 \rho_m = 4c^2 H^2/9\pi G = 9\hbar^2 G/64\pi c^2 r_e^6 = 7.385 \cdot 10^{-9} \text{ erg cm}^{-3} \qquad (31)$$

Planck equation causes the self-consistence of quantum physics models providing a conjunction of discrete corpuscular and continuous wave descriptions of a motion in microcosm. The self-consistency of the fundamental triad of quantum particle characteristics is determined by Planck constant. Substituting into Planck equation: $E = p^2/2m = h/t$ of the particle momentum defined by de Broglie equation: $p = hk = h/\lambda$ results in the quantum self-consistency condition: $m\lambda^2/t = h/2$. The analogous self-consistency condition for cosmological characteristics represents the quantum equation of the evolution (27). With the use of the relation $\rho_E = 3mc^2/4\pi r^3$ Eq. 27 can be transformed to:

$$m\tau/r^3 = 8H/9G = \hbar/2c^2 r_e^3 = 2.622 \cdot 10^{-11} \text{ g s cm}^{-3} \qquad (32)$$

This relation defines self-consistence of the matter distribution parameters in the process of the universe evolution.

Planck equation, defining a minimal action $\hbar/2$ is in fact a laconic formulation of the least action principle in the microcosm. Quantum equation of the universe evolution (27) having tight methodological unity with Planck equation also can be considered as brief formula of the least action principle in mega-world.

Odd feature of Standard model is the absence of fundamental electromagnetic field constants, even though all astrophysical data are exclusively derived from analysis of various forms of electromagnetic radiation. Quantum cosmology restores the key role of fundamental electromagnetic field constants in the universe evolution model.

### 4. Self-similarity of the universe and magic great numbers

Dimensional analysis in cosmology came to attention of astronomers after Dirac applied Eddington' "magic great numbers" to validate a new model of the universe evolution. Eddington noticed that dimensionless relation of electromagnetic and gravitational interactions between proton and electron $Ed_I = e^2/Gm_e m_p \simeq 2,3 \cdot 10^{39}$ is close to a value of the relation between estimated universe radius and classical electron radius: $Ed_{II} = R/r_e \simeq 3,6 \cdot 10^{40}$. It was also noticed that a relation between square root of estimated universe mass and proton mass is about the same value: $Ed_{III} = (M/m_p)^{1/2} \simeq 7,7 \cdot 10^{39}$. While Dirac' hypothesis has gone down in history, the unusual close coincidence of Eddington' magic numbers still defies all attempts of explanation.

Representation of the Hubble constant as a combination of fundamental constants (29) provides the relations of characteristic scales for mega- and micro-world, defined with a unique combination of fundamental constants. To gain an insight in the general mechanism of forming cosmological scales, we should consider the methods of introducing scales basing on universal constants. Thus, to use constants $\{c, G, e^2, m_e, m_p\}$ for a mass scales, one may employ: electron mass $m_e$ and proton mass $m_p$, the relation $m_G = (e^2/G)^{1/2}$ and scales of the type: $m_{Gi} \propto e^2/Gm_i$. In the same manner, one can use mass equivalent $m_{Ge}$ corresponding to electron electromagnetic energy $E_e = e^2/r_e$ considered as internal electron energy. Energy of the electron gravitational interaction with a relativistic object with mass $m_{Ge}$ at a distance $r_e$ is defined by the relation (see (6): $U_{Ge} = -3Gm_{Ge}m_e/2r_e$. In accordance with the virial theorem $E_e = -U_{Ge}/2$ and, therefore $e^2/r_e = 3Gm_{Ge}m_e/4r_e$. This relation defines the mass scale:

$$m_{Ge} = 4e^2 / 3Gm_e = 5.060 \cdot 10^{15} \text{ g} \tag{33}$$

This mass scale can be regarded as one of the estimates of the relation between forces of electromagnetic and gravitational interaction.

Using mass scales, one can derive corresponding length scales by applying the formulae: $l_i \propto e^2 / m_i c^2$ or $l_j \propto Gm_j / c^2$. This scale set includes, in particular, the classical electron radius:

$$r_e = e^2 / m_e c^2 = 2.818 \cdot 10^{-13} \text{ cm} \tag{34}$$

Time scales can be derived from length scales, using the formula $\tau_i \propto l_i / c$, for example:

$$\tau_e = e^2 / m_e c^3 = 9.400 \cdot 10^{-24} \text{ s} \tag{35}$$

This scale corresponds to the duration of light travel a distance equal to the classical electron radius (34).

Using the derived scales, one can develop multiple dimensionless relations and their functions. Some relations appear as indeed "great" numbers, for example those with denominator containing gravitational radiuses. Among them is the famous Eddington number $Ed_I = 2,3 \cdot 10^{39}$, corresponding to the relation of the classical electron radius to the proton gravitational radius.

Great cosmological numbers can be also derived as dimensionless constant combinations from relations: $M_H / m_i ; R_H / l_i ; T_H / t_i$. Dimensional analysis of fundamental constant groups, for example, $\{c, G, \hbar, e^2, m_e, m_p\}$ allows to produce scores of scales for the fundamental triad, in their turn permitting generation of hundreds dimensionless complexes. Including of the cosmological scales $\{T_H, R_H, M_H\}$ into the constant set will cause a many-fold increase of the number of dimensionless complexes. The hundreds dimensionless numbers thus produced can be used to find scores of triple great numbers coinciding in value to various degrees of accuracy. Of particular importance, however, is the scale triad in the system $\{c, e^2, m_e, G, H\}$ with relations not approximate, like in Eddingtons numbers, but *exact*:

$$K_T = M_H / m_{Ge} = R_H / r_e = T_H / t_e = \frac{3c}{2r_e H} = \frac{8r_e^2 c^3}{3\hbar G} = 8.105 \cdot 10^{40} \tag{36}$$

These equalities determine the dimensionless self-similarity criterion $K_T$ as a relation of characteristic mega- and micro-world scales which, being defined by a unique dimensionless value, are invariable during the universe evolution. Therefore, the universe in the course of evolution retains, despite monotone change of the scale-factor, the physical self-similarity with constant relation of mega- and micro world scales.

### 5. Fractal dimension of the large-scale structures in Metagalaxy: D = 2

During the last quarter of 20[th] century rapid progress of instruments for astronomical observations and achievements of computer technologies gave birth to new statistical techniques in investigation of "three-dimensional" distributions of matter in Metagalaxy. A conversion of two-dimensional projection of large-scale structure in Metagalaxy on the celestial sphere to a three-dimensional picture requires the estimation of third coordinates, using the galaxy redshifts with consequent calculation of their distances from Hubble law: $r = cz/H$. No more than a thousand galaxy redshifts were measured in the 1980s, more than a hundred thousand in the 1990s, and more than a million to date. It is well to bear in mind, however, that the three-dimensional picture of galaxy distribution obtained with the use of Hubble law is not the true three-dimensional section of the 4-dimensional space-time at some fixed moment. Hubble law only permits to estimate the distance to galaxies by line of sight for retrospective past moments of time. Thus, due to finite speed of light the distribution of galaxies at a distance, let us say, around 300 Mpc is now seen as it was almost a billion years ago. Therefore, the "three-dimensional" galaxy distribution examined with statistical techniques consists of a set of two-dimensional projections on celestial sphere of the true three-dimensional galaxy distributions but for a set of different consequent epochs.

The use of Hubble law for estimation of third coordinates in statistical analysis of galaxy distribution presumes employment of the relation: $r = c\tau$. With characteristic time from this relation: $\tau = r/c$ the Eq. 28 transforms to:

$$\rho_m = \frac{2cH}{3\pi G} r^{-1} = \frac{k_T}{c} \cdot r^{-1} = 1.878 \cdot 10^{-1} \cdot r^{-1} \text{ g cm}^{-3} \qquad (37)$$

For self-similar fractals, a special Hausdorf fractal dimension $D$ can be introduced by the relation: $m \propto r^D$, corresponding to mass density (see e.g. [1]): $\rho_m \propto r^{-(3-D)}$. A comparison of this relation with Eq. 37 shows that Eq. 37 describes the mass density of cosmic structures with fractal dimension:

$$D = 2 \qquad (38)$$

Therefore, statistical methods of galaxy distribution analysis, employing Hubble law to define third coordinates, should disclose the fractal dimension of distributions $D = 2$.

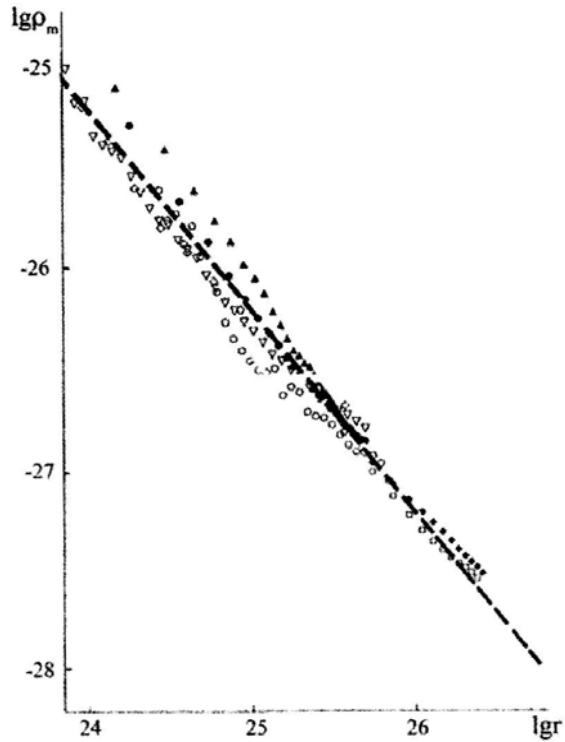

*Fig. 1 Observational data from different galaxy catalogues [7] compared with theoretical equation (37) (dotted line).*

Fig. 1 gives comparison of formula (37) with results of statistical analysis of galaxy distributions [7]. Different marks at figure correspond to analyses with data from different galaxy catalogues. As the Fig. 1 demonstrates, theoretical formula (37) (dotted line) agrees satisfactorily with observational data at least up to a distance scale around 300 Mpc.

## 6. Universe evolution as the global energy source

Standard cosmological model cannot explain from where the present immense mass of the universe appeared if the initial Planck mass of the newborn universe was infinitesimal: $m_{Pl} = 2.2 \cdot 10^{-5}$ g. Increasing radius of the expanding universe with invariable mass should result in gradually decreasing universe gravitational energy. In the relativistic ideology with variable number of particles and where the energy conservation law includes energy equivalent of potential mass change, decreasing gravitational energy in an isolated system can be only compensated by the increase of internal energy and mass. That is, at least in part, by generation of the new matter. Characteristics of possible universe mass growth can be estimated from analysis of the relativistic relation (7) for an isolated system. From this relation and the evolution description in the form (4, 22) it follows that the relation defining universe mass growth is:

$$M = \left(8c^6/3HG^2\right)^{1/2} \tau^{1/2} \qquad (39)$$

Average rate of the universe mass growth can be estimated with the use of cosmological scales (24 – 26): $Q_M = M_H/T_H = 4c^3/3G = 5 \cdot 10^{38}$ g s$^{-1}$. To gain a visual impression of the universe mass growth, one may use an estimate of the mass growth rate in a unit of Metagalaxy volume: $Q_M/V = 8H^3/27\pi G = 1.1 \cdot 10^{-47}$ g s cm$^{-3}$. This mass growth rate means, for instance, that in the whole volume of the Earth during all its history could appear no more than $2 \cdot 10^{-3}$ g of hydrogen, not enough to fill a child balloon. The relative universe mass growth also seems insignificant: $\delta_M = Q_M/M_H = 2H/3 = 1.3 \cdot 10^{-18}$ s$^{-1}$. Yet in the whole Metagalaxy this mass growth means the birth of new cosmic objects with the total mass of more than $10^5$ solar masses, i.e. of the same order as masses of globular star cluster or a dwarf galaxy, emerging every second.

The estimated characteristics of the universe mass growth discussed above by no means suggest uniform matter synthesis across the Metagalaxy. It seems rather that such high-energy processes occurred and, most likely, still proceed in relatively few centers, like quasars or active nuclei of massive galaxies.

The universe evolution results in the universe mass growth and in accordance with relativistic ideology in a growth of the universe internal energy. Total power of this global process can be estimated as current value of the time-derivative of (39):

$$W_p = \left|d/d\tau(c^2 M)\right|_{\tau_p=T_H} = \left(2c^{10}/3HG^2\right)^{1/2} T_H^{-1/2} = 2c^5/3G = 2.420 \cdot 10^{52} \text{ W} \qquad (40)$$

Non-stationary state of the universe appears as a global source of energy, and we could try to identify "evolutional" energy effects, entering cosmological *terra incognita*. Probably major part of the evolutional energy consumed by the processes of the new matter synthesis, but, alas, we know nothing about these processes. We may suggest that some fraction of the evolutional energy is absorbed by cosmic objects, in particular, by planets. Local non-stationary state of the space-time may be allegorically described by the local cosmological space "expansion" and by apparent "growth" of massive cosmic bodies with consequent their gravitational energy decrease [14]. For an isolated massive cosmic body with the constant total energy a decrease of potential gravitational energy during its cosmological "expansion" ought to result in the compensating growth of its internal energy. Cosmological "increase" of the planet radius in accordance with Hubble law: $r(t) = r_0(1 + Ht)$ must lead to decrease of gravitational energy: $U_G = -Gm^2/r_0(1 + Ht)$, initiating a compensating energy transfer of the initial gravitational energy into internal thermodynamic energy.

Defining the energy flow from the planet interior as a certain fraction of the current momentary cosmological change of the planet potential gravitational energy we can get:

$$L \propto \left|dU_G/dt\right|_{t=t_p} = \left|d/dt[-Gm^2/r_0(1 + Ht)]\right|_{t=t_p} = HGm^2/(1 + Ht_p)r_p \propto -HU_G \qquad (41)$$

Here $t_p$ is the planet Newtonian age. In transformations leading to Eq. 41 the relation: $r_p = r_0(1 + Ht_p)$ is used. The Eq. 41 corresponds to virial theorem. The heat flow from planet interior is proportional to planet internal energy that in its turn in accordance with virial theorem is proportional, as demonstrates Eq. 41, to planet gravitational energy: $L \propto -U_G$.

To estimate the proportionality coefficient in Eq. 41 the investigation of the heat flow from the Earth interior can be used. In the last quarter of the 20$^{th}$ century thousands of heat flow measurements were performed in different regions of our planet, and the Earth heat flow is reliable estimated as $(4.2 - 4.5) \cdot 10^{13}$ W (see e.g. [17]). It appears that the energy of radioactive minerals decay is not enough to explain the Earth internal heat flow (see e.g. review in [17]). Several independent studies showed that at present the generation of radiogenic energy in the Earth does not exceed $1.3 \cdot 10^{13}$ W (out of which $0.9 \cdot 10^{13}$ W in the earth crust and $0.4 \cdot 10^{13}$ W in the earth mantle) that totals only 30 % of the entire heat flow. Therefore, the gravitational component of the energy flow not related to radio-chemical processes is about $L = (2.9 - 3.2) \cdot 10^{13}$ W. For theoretical value of Hubble constant (29) corresponding proportionality coefficient in Eq. 41 is $(4.15 \pm 0.2) \cdot 10^{-2}$ and Eq. 41 becomes:

$$L = -(4.15 \pm 0.2) \cdot 10^{-2} \cdot HU_G = (4.15 \pm 0.2) \cdot 10^{-2} \cdot HG \frac{m^2}{r} \quad \text{erg s}^{-1} \tag{42}$$

To compare Eq. 42 with observations it is possible to use key energy parameters of the Earth and outer planets (see e.g. Table 4.3 in [14]). Fig. 2 represents in decimal logarithmic coordinates the comparison of the formula (42) (dotted line) with the estimations of heat flows based on the astrophysical data.

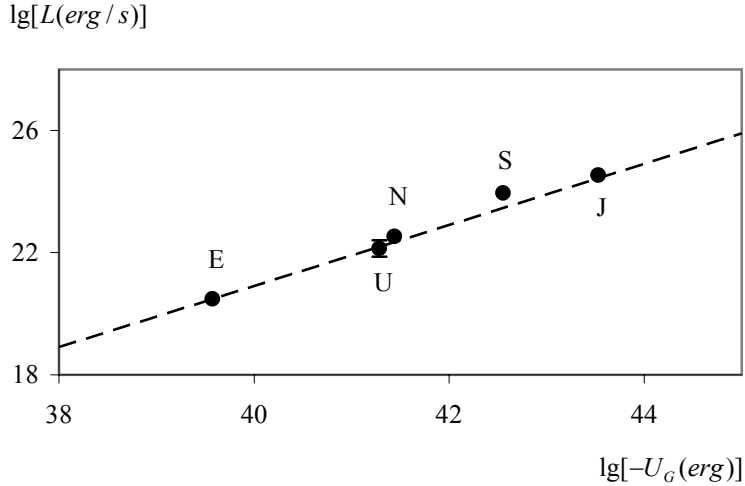

*Fig. 2* *The observational estimations of surface heat flows (filled circles) compared with Eq. 41 (dotted line). E – Earth, U – Uranus, N – Neptune, S – Saturn, J – Jupiter.*

Fig. 2 demonstrates a satisfactory coincidence of the formulae (41, 42) with observational data, and, in particular, Eq. 42 can be used for estimation of heat flows from the outer planet interiors with the error that does not exceed one standard deviation for the observational data. An important result of our analysis is the independence of the planet heat flows on the chemical and structural characteristics of planet interiors. Energy balances of planets governed only by their gravitational energy and by Hubble constant, defining global rate of the universe evolution. Quite good agreement of theoretical analysis of the planet heat balances, based on the estimations of the global cosmological evolution effects, with observational data may be regarded as one of the arguments in favor of the assumption about existence of the evolutional energy effects caused by the local non-stationary space-time state.

### 7. Large-scale structure parameters in Metagalaxy

Quantum cosmology is attractive not only by its analysis of the unity of physical laws in the mega- and micro world. An advantage of quantum cosmological models is the effective description of the large-scale structure parameters in Metagalaxy. Observational data discussed in this article suggest that all physical processes in a non-stationary universe evolve in the cosmologically decelerating physical time and just this time should be used in mathematical models of the large-scale structure of Metagalaxy. This assumption raises a question: how one can be certain of the physical time advantages in astrophysics if there are no methods of direct estimation of the time intervals between astrophysical events?

Mathematical models in cosmology can be formulated as the general relations:

$$F(t; x_i, ...) = 0 \quad \Phi(\tau; x_i, ...) = 0 \tag{43}$$

Here $x_i$ stands for the observable parameters of cosmic structures. To exclude time from these mathematical models, cosmology uses the relation between luminosity distance $r_L$, estimated from apparent magnitudes, and the redshift, with an assumption of the speed of light constancy: $t = r_L(z;...)/c$ and $\tau = r_L(z;...)/c$. Using these relations, we can obtain from Eq. 43:

$$x_i = f_i(z;...) \quad x_i = \varphi_i(z;...) \tag{44}$$

Comparing these relations with observational data, we can judge the correctness and usefulness of the concept of cosmological deceleration of the course of physical time.

Solutions of the quantum cosmological model equations (11, 12) are useful in defining relations for cosmological distances and redshifts. For a frame of reference with zero-time reference point at the emission moment Eq. 14 is valid, and the following initial conditions for Eq. (11, 12) can be used: $a = 1+z$ and $a(\tau = 0) = a_0 = 1$. In this case Eq. 16 and formula $\tau = r_L/c$ define the following relations between redshift and luminosity distance $r_L$:

$$z = \left(1 + 2H \frac{r_L}{c}\right)^{1/2} - 1 \tag{45}$$

$$r_L = \frac{c}{2H}[(1+z)^2 - 1] \tag{46}$$

Using the dependence of luminosity distance $r_L$ (Mpc) on distance modulus $\mu = m - M$, Eq. 46 can be rewritten in the form convenient for comparison with astrophysical observations:

$$\mu_T = 5\lg\left\{\frac{c}{2H}[(z+1)^2 - 1]\right\} + 25 \tag{47}$$

The corresponding formula for Newtonian time derived from Hubble law (18) is:

$$\mu_H = 5\lg\left(\frac{cz}{H}\right) + 25 \tag{48}$$

Formulae (47) and (48) are a good example of the relations (44).

Observational data for distant bright quasars at $z > 2.5$ and $\langle M \rangle_g = -28.2^m \pm 0.3^m$ (see e.g. Table A5 in [14]) can be combined with data for supernovae SNe Ia (see e.g. Table A3 in [14]) to test Eqs. 47, 48 with theoretical value of Hubble constant (29) in the wide redshift range: $z = 0 - 4.65$. Spectral bands of supernovae apparent magnitudes data (B: 0.45 mcm) and quasar apparent magnitudes data (g: 0.47 mcm) are almost the same and for distance modulus the following relations: $\mu = m_B + 19.5$ (supernovae) and $\mu = m_g + 28.2$ (quasars at $z > 2.5$) can be used.

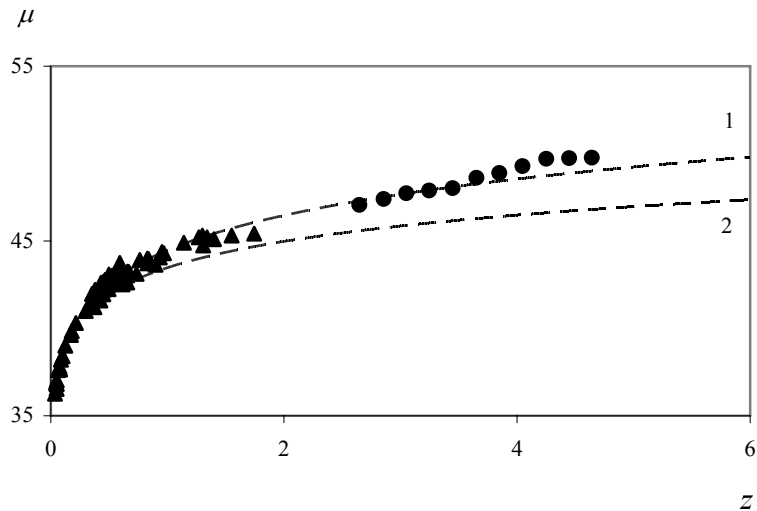

***Fig. 3** The observational data (filled triangles for SNe Ia, filled circles for quasars at $z > 2.5$) compared with Eq. 47 (dotted curve 1) and Eq. 48 (dotted curve 2).*

Fig. 3 demonstrates that Eq. 47 matches observational data much better than Eq. 48. A difference in apparent magnitude estimates by Eqs. 47 and 48: $m_T - m_H = 5\lg(1 + z/2)$ becomes substantial at $z > 2$, for example at $z = 4$ the relative divergence comes up to 11 %.

In quantum cosmology a doctrine of "expanding" space of the universe: $\Delta l(\tau) = a(\tau)\Delta l_0$ with monotone increasing scale-factor $a(\tau)$ together with a condition of the constancy of the speed of light: $\Delta l = c\Delta\tau; \Delta l_0 = c\Delta\tau_0; c = const$ lead to the relation: $\Delta\tau(\tau) = a(\tau)\Delta\tau_0$ with unlimited extent of time intervals. Using the Eq. 14 to transform this relation, one can get an equation defining cosmological growth of time intervals:

$$\tau = \tau_0(1+z) \qquad (49)$$

Here $\tau_0$ is the time interval at $z = 0$. Eq. 49 suggests cosmological growth for both microscopic time intervals like photon periods, and quite bigger macroscopic time intervals [13, 14].

Recent studies of supernovae SNe Ia (with $z$ up to 0,85) discovered an expansion of supernova light-curves (time-dependences of luminosity) with the growth of redshifts [3]. Fig. 4 with plotted relative luminosity periods $\tau_k$ of SNe Ia corresponding to observational data (see e.g. Table A3, column 7 in [14]) demonstrates satisfactory agreement with Eq. 49 (dotted line), attesting to appreciably increase of the supernova luminosity times with the growth of distances to them. The value $\tau_0 = 1 \pm 0.14$ in Eq. 49 is the SN initial relative luminosity time derived from the observational data at $z \to 0$.

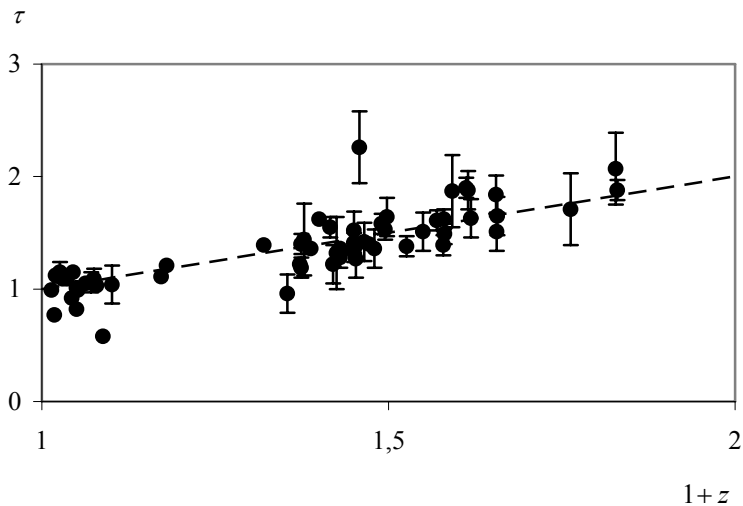

***Fig. 4** The agreement of Eq. 49 (dotted line) with observational data for supernovae SNe Ia.*

A concept of cosmological deceleration of the course of physical time explains the phenomenon of the supernova luminosity time growth in the same way as the increase of photon period in the redshift phenomenon: all process durations in the past seem longer in our epoch due to cosmological deceleration of time and the increase of time standards. But the increase of SN luminosity times differs fundamentally from the redshift phenomenon. The photon period $\Delta\tau_{ph}$ (around $2\cdot 10^{-15}$ s) is a typical microscopic quantum parameter, and the redshift can be explained, in accordance with the quantum relation $\Delta\tau_{ph} = \lambda/c$, by alone photon wavelength growth in the expanding universe space. On the other hand, the SN luminosity period is usually around four weeks i.e. about $10^6$ s, being by 21 orders of magnitude over the photon period and belonging not to micro- but to macro parameters. No reasonable characteristic length could be found for this macro parameter to explain its growth by the expanding universe space. The phenomenon of increasing SN luminosity times with the growth of the redshifts provides an impressive evidence of the macroscopic time interval enlargement due to cosmological deceleration of the course of physical time.

A constancy of the speed of light enables to describe any length with the relations: $d_L = c\tau$ and $d_{L0} = c\tau_0$. Using these relations, Eq. 49 can be transformed to:

$$d_L = d_{L0}(1+z) \tag{50}$$

This relation allows to suggest not only cosmological growth of microscopic parameters like the photon wavelength, but the increase of apparent macroscopic dimensions of cosmic structures like angular diameters when distances between observer and an object grow up. Apparent size of a cosmic object, for instance the luminosity radius $R_L$, will increase according to (50) with the growth of a distance from the observer. Using Eqs. 50 and 46 one can derive the following relation for the angular dimension:

$$\theta \simeq tg\theta = R_L/r_L \propto \frac{1+z}{(1+z)^2 - 1} \tag{51}$$

To test this relation observational data published in [2] can be used. The authors collected and analyzed statistics of 25 elliptic galaxies (Table A4 in [14]) at various distances at $z = 0,00317 \div 1,175$. Fig. 5 illustrates a comparison of Eq. 51 with published observational data [2] in decimal logarithmic coordinates, using angular galactic radii $\theta$ in radians. As one may see there is a satisfactory agreement of the observational data with theoretical estimation of Eq. 51.

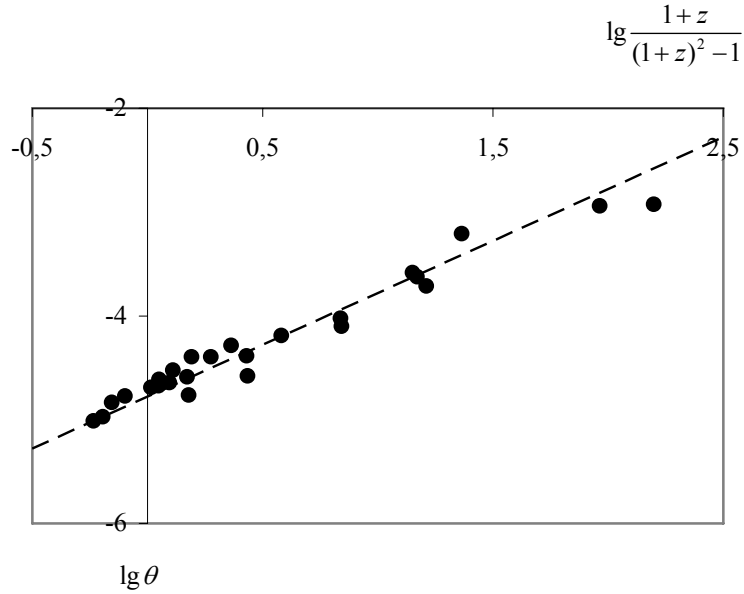

*Fig. 5 Elliptic galaxy observational data (filled circles) compared with Eq. 51 (dotted line).*

Slow increase of apparent magnitudes with the growth of redshifts and a complex form of quasar parameters domain in $\{lg(cz) - m\}$ diagram lead to the widespread opinion that Hubble law in the forms (18) or (21) cannot be used in analysis of quasar observational data. However, Hubble law appears a quite effective tool if additionally the influences of quasar dimension and structure are taken into account. It helps to avoid the overestimated quasar luminosities and reveals statistical parameter relations hidden in the exotic form of quasar domain in $\{lg(cz) - m\}$ diagram.

It can be suggested that the estimation of quasar luminosity depends on its volume and apparent angular diameter, since the main source of quasar luminosity usually is linear structure of the quasar plasma jets. Therefore luminosity should be directly proportional to the quasar volume $V \sim d_L^3$ and inversely proportional to its angular diameter: $L = L_0 d_L^3 / \theta$. This relation after use of Eqs. 50, 51 transforms to:

$$L = L_0 d_L^3 / \theta = L_0 (1+z)^2 [(1+z)^2 - 1] \tag{52}$$

Here $L_0$ is the quasar luminosity at a standard distance of 10 pc from the observer, i.e. at $z \simeq 0$. The relation between luminosity and observed radiation flux is described by: $F = L/r_L^2$. Using Eqs. 46 and 52, the relation for the quasar flux can be transformed to:

$$F = L/r_L^2 = L_0 \frac{4H^2(1+z)^2[(1+z)^2 - 1]}{c^2[(1+z)^2 - 1]^2} \qquad (53)$$

The formula for quasar apparent magnitudes can be derived from the definitions of apparent and absolute magnitudes: $m = M - 2,5\lg(F/L_0) + 25$. In this formula luminosity distance is estimated in Mpc. After introduction of Eq. 53 and the value: $\lg(4H^2/c^2) = -6.785$ (with theoretical Hubble constant (29)), this formula for quasar apparent magnitudes becomes:

$$m = M + 5\lg[(1+z)^2 - 1] - 2,5\lg\{(1+z)^2[(1+z)^2 - 1]\} + 41.96 \qquad (54)$$

The second term in this formula accounts for the decrease as $1/r_L^2$ of the radiation flux. The third term describes the influence of quasar volume and plasma jet size on its apparent magnitude.

A possible approach to compare Eq. 54 with observational data for quasars is to use the calculated from catalog SDSS DR6 (Sloan Digital Sky Survey. Data Release 6) [11] average magnitudes $\langle m_r \rangle$ (r: 0.62 mcm), $\langle m_g \rangle$ (g: 0.47 mcm) and $\langle z \rangle$ for intervals, for example: $\Delta z = 0.1$ (see e.g. Table A5 in [14]). A differences between $\langle m_r \rangle$ and $\langle m_g \rangle$ is less than their standard deviations at least up to $\langle z \rangle = 4$ and due to it as average apparent magnitudes the values of $\langle m \rangle = (\langle m_r \rangle + \langle m_g \rangle)/2$ can be used. Fig. 6 illustrates a comparison of the averaged observational data (filled circles) with the Eq. 54 (dotted curve 1). Estimations yield the average absolute quasar magnitude $\langle M \rangle = -23^m \pm 0.2^m$ for quasars with redshifts in the range $\langle z \rangle = 0 \div 2.5$, which is less than often published estimates and corresponds to absolute magnitudes of large galaxies.

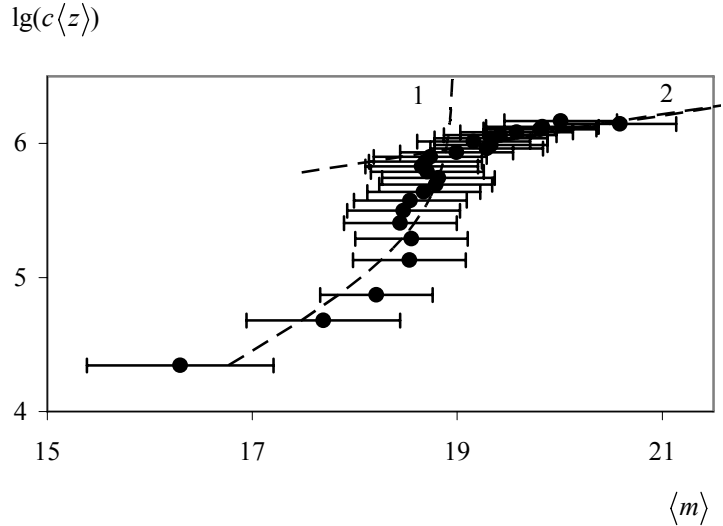

***Fig. 6*** *Quasar observational data (filled circles, corresponding to average apparent magnitudes, plotted with interval $\Delta z = 0.2$), compared with Eq. 54 (dotted curves 1) and Eq. 55 (dotted curve 2).*

At $z > 2.5$ quasar dimensions and structure don't influence the estimates of their apparent magnitudes and quasars can be considered as radiation point-sources. The magnitude-redshift dependence at $z > 2.5$ is described by Eq. 54 without the third term:

$$m = M + 5\lg[(1+z)^2 - 1] + 41.96 \tag{55}$$

Selection effects are of crucial importance for large distances and at $z > 2.5$ only high-luminosity quasars with absolute magnitudes $\langle M \rangle = -28.2^m \pm 0.3^m$ (dotted curve 2 in Fig. 6) are observable. Eqs. 54, 55 enable one to estimate quasar luminosities, using the relation: $\lg L = \lg L_\odot + (M_\odot - M)/2.5$. Analysis of the catalog data [31] indicates that all quasars fall into the two main groups:
- quasars at $z < 2.5$ (more than 90 % of data in [31]) with absolute magnitudes in the range: $M = -22.6^m \div -23.5^m$ with the mean value $\langle M \rangle = -23^m \pm 0.2^m$, corresponding to luminosities $L = (3-8) \cdot 10^{37}$ W, typical for large galaxies
- distant bright quasars at $z > 2,5$ with absolute magnitudes in the range: $M = -27.7^m \div -29.8^m$ with the mean values: $\langle M \rangle_g = -28.2^m \pm 0.3^m$ and $\langle M \rangle_r = -29.1^m \pm 0.4^m$, corresponding to luminosities $L = (4-27) \cdot 10^{39}$ W.

**8. Thermal asymmetry of the Cosmic Microwave Background radiation**

The study of Cosmic Microwave Background radiation (CMB hereafter) has become one of the cornerstones of current observational cosmology, as well as a major data pool for astrophysics in order to test cosmological models and fundamental physics theories. After Arno Penzias and Robert Woodrow Wilson discovered in 1965 this radiation, filling the entire universe, many peculiar CMB characteristics were investigated: precise measurment of the CMB average temperature, the thermal black body form of the CMB spectrum, the CMB power spectrum temperature anisotropy in terms of the angular scale (dipole and multipole moments), CMB polarization, and it seems that this inventory is not finished. As the precision and angular resolution of CMB data sets get better, new challenges arise.

There are two basically different approaches to the nature of the CMB. According to standard cosmology CMB is a snapshot of the early universe stage some 400,000 years after the Big Bang, when the temperature dropped enough (down to 4000 - 3000 K) to allow electrons and protons to form hydrogen atoms, thus making the early universe transparent to radiation. Since this period of recombination or "decoupling" in primeval electron-proton plasma the temperature of the CMB radiation has dropped by a factor of roughly 1100 (down to 2.7 K) due to the expansion of the universe.

After the observational detection of an average cosmic bolometric temperature of 2.3 K by Andrew McKeller in 1941, the second aproach to the CMB origin was discussed. According, for example, to Fred Hoyle the CMB could be a remnant of the evolution of stars. Its energy density is equal to the energy released by the nuclear reactions in stars of all generations. The optical photons radiated by stars could be "thermalized" by scattering and gravitational deflections in intergalactic electron-proton plasma. Observations of the WMAP (Wilkinson Microwave Anisotropy Probe, 2001 - 2003) satellite provided evidence that the intergalactic medium was ionized at very early times, at a redshift of larger than 17. Neverthless during the 1990s the consensus was established that the CMB is a remnant of the Big Bang.

In spite of all differences in these two models of CMB origin, in both models there is a common element - the electron-proton plasma, playing a role of the radiation-transforming medium. One of the possible ways to estimate CMB parameters is to consider the CMB emergence as the scale-invariant radiation transformation in the electron-proton plasma (Taganov, 2008 [14, 15]). To perform the scale-analyzis one may employ the power-law distributions of certain quantities and a set of characteristic scales. Analizing CMB radiation it is reasonable to use the general relation for cosmological energy densities $\rho_E$ in the form of power-law distribution (9, 27): $\rho_i = k_T \tau_i^{-1}$. This power-law energy density-time dependence allows to get for two scale-invariant cosmic systems the following relation:

$$\rho_i / \langle \rho_E \rangle = \tau_i / \langle \tau \rangle \tag{56}$$

The average energy density of the universe $\langle\rho_E\rangle = 7.385 \cdot 10^{-9}$ erg cm$^{-3}$ defined by Eq.31. Scaling properties of electrons, protons and nucleon-electron structures in cosmic plasma with fundamental constants $\{m_e, e, \hbar\}$ can be described by three sets of quantum scales:

1. Compton electron wavelength: $\lambda_{Ce} = \hbar/m_e c$ and the time scale: $\tau_e = \lambda_{Ce}/c = \hbar/m_e c^2 = 1.288 \cdot 10^{-21}$ s.
2. Compton proton wavelength: $\lambda_{Cp} = \hbar/m_p c$ and the time scale: $\tau_p = \lambda_{Cp}/c = \hbar/m_p c^2 = 7.016 \cdot 10^{-25}$ s.
3. Radius of the first Bohr orbit: $a_B = \hbar^2/m_e c^2$ and the time scale: $\tau_{ep} = \hbar^3/m_e e^4 = 2.421 \cdot 10^{-17}$ s.

After substitution of time scales: $\tau_i = \tau_e = \hbar/m_e c^2$ and $\langle\tau\rangle = \tau_{ep} = \hbar^3/m_e e^4$ into Eq. 56 it takes the form:

$$\rho_{HMB}/\langle\rho_E\rangle = \tau_e/\tau_{ep} = e^4/c^2\hbar^2 = 5.320 \cdot 10^{-5} \tag{57}$$

Here $\rho_{HMB}$ is the energy density of the *high-frequency microwave background* (HMB). The use of the average energy density estimation (31) allows deriving from Eq. 57 a relation for the HMB energy density:

$$\rho_{HMB} = \langle\rho_E\rangle e^4/c^2\hbar^2 = 3.929 \cdot 10^{-13} \text{ erg cm}^{-3} \tag{58}$$

The HMB temperature can be defined from this energy density with the use of Stephan-Boltzmann constant ($\sigma^* = 4\sigma/c = 7.566 \cdot 10^{-15}$ erg cm$^{-3}$ K$^{-4}$; $\sigma = 5.670 \cdot 10^{-8}$ W m$^{-2}$ K$^{-4}$):

$$T_{HMB} = \left(c\rho_{HMB}/4\sigma\right)^{1/4} = \left(\rho_{HMB}/\sigma^*\right)^{1/4} = 2.684 \text{ K} \tag{59}$$

This scaling temperature estimation is surprisingly accurate since it differs by less than 2 % from, say, the CMB temperature: $T_{CMB} = 2.728 \pm 0.004$ K, precisely measured by the COBE (NASA Cosmic Background Explorer satellite, 1989 - 1996).

The use of time scales $\tau_i = \tau_p = \hbar/m_p c^2$ and $\langle\tau\rangle = \tau_{ep} = \hbar^3/m_e e^4$ allows to estimate from (56) the energy density of the *low-frequency microwave background* (LMB):

$$\rho_{LMB} = \langle\rho_E\rangle \tau_p/\tau_{ep} = \langle\rho_E\rangle e^4 m_e/c^2\hbar^2 m_p = 2.139 \cdot 10^{-16} \text{ erg cm}^{-3} \tag{60}$$

Corresponding temperature of LMB radiation one can evaluate similar to Eq. 59:

$$T_{LMB} = \left(\rho_{LMB}/\sigma^*\right)^{1/4} = 0.41 \text{ K} \tag{61}$$

Comparing HMB and LMB radiations it is helpful to note that the maximum HMB intensity corresponds to: $\nu_{HMB}^{max} = 159$ GHz and $\lambda_{HMB}^{max} = 0.189$ cm, while the maximum LMB intensity conforms to: $\nu_{LMB}^{max} = 24.1$ GHz and $\lambda_{LMB}^{max} = 1.243$ cm. Maximum intensities ratio is: $I_{LMB}^{max}/I_{HMB}^{max} = (T_{LMB}/T_{HMB})^3 = 3.5 \cdot 10^{-3}$. HMB and LMB radiations compose a joint spectrum of cosmic microwave background radiation (CMB) with average temperature slightly higher than HMB temperature (59): $T_{CMB} - T_{HMB} = 0.35$ mK.

The presence of two sources of microwave radiation with different emission temperatures causes the CMB *spectral temperature asymmetry*. Blackbody LMB spectrum can be represented with the use of Planck function $p(\nu; T) = 1/[\exp(h\nu/k_B T) - 1]$ as:

$$I_{LMB}(\nu; T) = 2h\nu^3/c^2 \, p(\nu; T_{LMB}) \tag{62}$$

Total blackbody CMB spectrum is described by the relation:

$$I(\nu;T) = 2h\nu^3 / c^2 [p(\nu;T_{LMB}) + p(\nu;T_{HMB})] \quad (63)$$

From this equation the formula for CMB spectral temperature can be derived:

$$T_{CMB}(\nu) = h\nu / k_B \ln\{1 + 1/[p(\nu;T_{LMB}) + p(\nu;T_{HMB})]\} \quad (64)$$

This relation estimates CMB spectral temperature corresponding to spectral intensity in narrow frequency interval. Averaging Eq. 64 in a wide frequency interval one can evaluate the CMB mean spectral temperature $\langle T_{CMB} \rangle$. The excess of predicted total CMB intensity (63) over intensity of dominant HMB radiation with temperature (59): $T_{HMB} = 2.684$ K is substantial only in the low-frequency range (curve 1 in Fig. 7). For example, the excess of predicted CMB intensity (63) over intensity of HMB (59) with temperature $T_{HMB} = 2.684$ K grows with the decrease of frequency: from 1 % at 25 GHz to 10 % at 5 GHz.

With the use of integral mean spectral temperature:

$$\langle T_{CMB} \rangle_{\nu_1}^{\nu_2} = \int_{\nu_1}^{\nu_2} T_{CMB}(\nu) d\nu / (\nu_2 - \nu_1) \quad (65)$$

the thermal asymmetry of the CMB for frequency range $\nu_1 \div \nu_2$ can be estimated as:

$$A_{\nu_1}^{\nu_2} = \langle T_{CMB} \rangle_{\nu_1}^{\nu_2} / T_{HMB} \quad (66)$$

For example, for frequency range $10.9 \div 16.3$ GHz the CMB thermal asymmetry is:

$$A_{10.9}^{16.3} = \langle T_{CMB} \rangle_{10.9}^{16.3} / T_{HMB} \simeq T_{CMB}(10.9) + T_{CMB}(16.3) / 2T_{HMB} = 1.063 \quad (67)$$

Predicted CMB *thermal asymmetry* or low-frequency CMB "overheat" does not contradict precise measurements of FIRAS (the Far-Infrared Absolute Spectrophotometer) aboard COBE satellite that have demonstrated deviations from the CMB blackbody spectrum form not more than $5 \cdot 10^{-5}$, since these measurements only made in the high-frequency range 60 - 600 GHz. In this high-frequency range the predicted share of LMB intensity in the total CMB intensity is less than $10^{-5}$.

Probably CMB thermal asymmetry and low-frequency CMB overheat already has been registered in COSMOSOMAS experiment (COSMOlogical Structures On Medium Angular Scales; Teide Observatory, Tenerife, Spain). COSMOSOMAS research team recently reported that unresolved extragalactic sources are found to be dominant foreground at 11 GHz as a signal detectable in the frequency range 11 - 33 GHz with amplitude of order $3 - 6 \mu K$ at 11 GHz [5]. The presence of CMB component with suggested temperature 2.728 K detected as the signal with amplitude $27 \pm 2 \mu K$ in the COSMOSOMAS channels. With this average CMB amplitude anomalous signal from unresolved extragalactic sources corresponds to predicted CMB low-frequency overheat of order 2.9 - 3.0 K.

Table 6 in [5] represents the temperature cross correlations excesses (E) between the COSMOSOMAS and WMAP_K map after contributions from CMB and Galactic emission are accounted. It appears that the temperature cross correlations excesses between the COSMOSOMAS and WMAP_K (Wilkinson Microwave Anisotropy Probe, 2001 - 2003) map, attributed to signals in common other than CMB with temperature 2.728 K, do not depend on the galactic latitude and decrease with increasing frequency.

The temperature cross correlations excess is proportional to the difference between the CMB spectral temperature and dominant HMB spectral temperature: $E \propto (T_{CMB} - T_{HMB})$. To find the proportionality coefficient in this relation it is reasonable to use most accurate WMAP_K measurements at frequency 23 GHz. The correlation between measured in [5] temperature cross correlations excesses at 23 GHz and theoretical estimation from Eq. 64 is: $(T_{CMB} - T_{HMB}) \simeq 0.0073E$. The use of this correlation allows evaluating of the CMB temperatures corresponding to different cross correlations excesses.

Fig. 7 demonstrates CMB spectral temperature-frequency dependence (curve 1 corresponds to Eq. 64), which shows bigger low-frequency CMB temperature than high-frequency one. Observational COSMOSOMAS data conform quite well to the theoretical prediction of Eq. 64. The evaluation of the CMB thermal asymmetry (66) for frequency range $10.9 \div 16.3$ GHz, corresponding to COSMOSOMAS channels, using observational data with the CMB high-frequency temperature: $T_{CMB} = T_{HMB} = 2.684K$, yields $A_{10.9}^{16.3} = 1.064$ that almost exactly coincides with theoretical estimation (67).

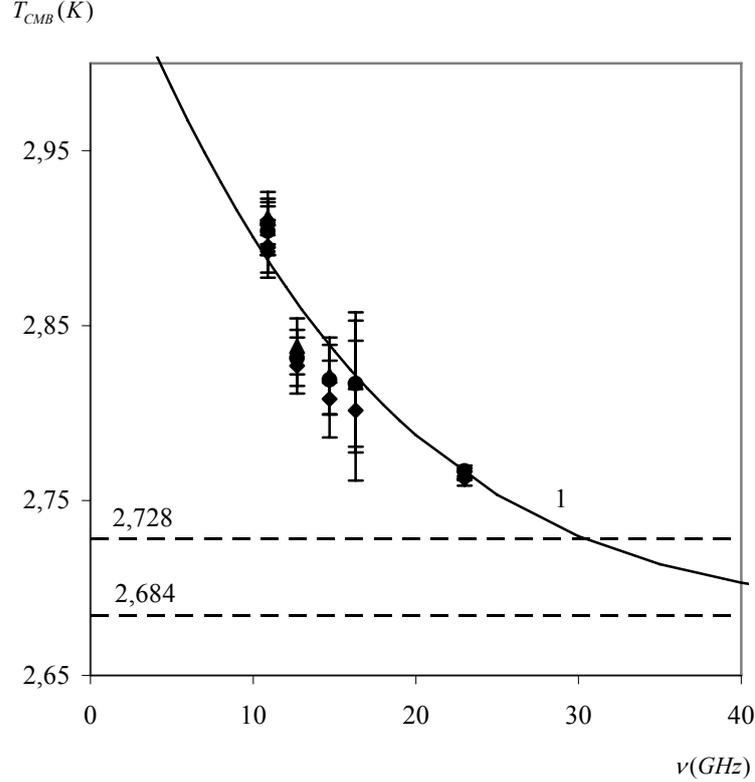

*Fig. 7 Predicted (curve 1) and observed CMB spectral temperatures in low-frequency channels [5]. Filled circles for $|b| > 30^o$, filled triangles for $|b| > 40^o$, filled diamonds for $|b| > 50^o$.*

A dust correlated emission is detected in all COSMOSOMAS channels at $|b| > 30^o$. The amplitude of the signal ranges from $10-12\mu K$ at 11 GHz down to $4-7\mu K$ in the 12 - 17 GHz and $2-3\mu K$ at 23 GHz. A considerable part of this correlated signal at 11 GHz comes from regions of high dust emission where free-free emission is not well traced by the $H\alpha$ template due to extinction. After masking those regions the remaining anomalous signal still detectable in the frequency range 11 - 41 GHz with amplitude of order $3-6\mu K$ at 11 GHz. This anomalous signal shows a clear flattening in the frequency range 11 - 22 GHz, which is not compatible with the classical spectral index of synchrotron emission.

Fig. 8 reproduce observational data of Table 10 ($|b| > 30^o$) in [5] with flux densities of anomalous signal in units Jy/sr (Fig 17 in [5]) for correlation between COSMOSOMAS, WMAP channels and the DIRBE08 map after masking some localized regions. Authors of [5] found that the single Gaussian law with maximum at $21.7^{+3.8}_{-3.7}$ GHz and $\sigma = 15.8^{+4}_{-3.4}$ GHz mimics this anomalous signal. This approximation corresponds well to predicted LMB maximum at 24.1 GHz. However, even better than to Gaussian law these observational data correspond to predicted LHB spectrum: Eq. 62 (curve 1 in Fig. 8).

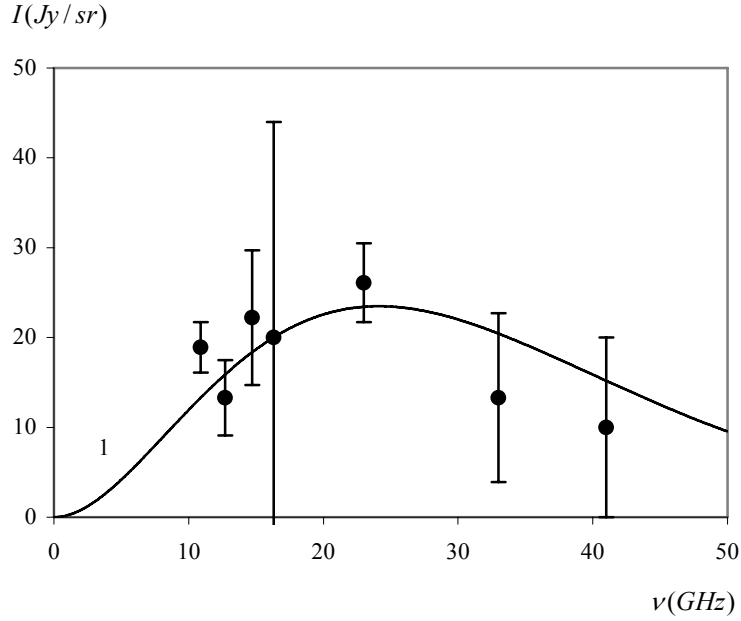

*Fig. 8 Correlation between COSMOSOMAS, WMAP channels and the DIRBE08 map in flux density units (Jy/sr) for $|b| > 30^o$ (Fig. 17 in [5]). Overplotted curve 1 is the predicted LMB intensities (Eq. 62).*

Quantum scaling methodology used in this paper predicts only basic CMB characteristics: LMB and HMB energy densities and emission temperatures, but cannot reveal the physical processes of the CMB origin. Authors of [5] suppose that the registered by COSMOSOMAS low-frequency anomalous signal could be explained by a combination of free-free emission and spinning dust models with a flux density peaking around 20 GHz. However, spinning dust models predict fast emission decrease with the decline of frequency in the range 10 - 5 GHz. Hence the remaining free-free emission with spectral index $\alpha = -0.1 \div -0.3$ would give almost constant intensity in the frequency range <5GHz. On the contrary, predicted CMB thermal asymmetry causes a stable intensity decrease at frequencies <10 GHz (curve 1 in Fig. 2). Therefore, a crucial test for predicted phenomenon of CMB thermal asymmetry is the precise CMB measurement in low frequency range < 10 GHz.

**Conclusion**

In 1927 G. Gamov, D. Ivanenko and L. Landau proposed a convenient classification of physical theories by means of three fundamental constants: $G, c, h$, used by Planck early in the 20[th] century to introduce his "natural" primary units. For example, the classical mechanics was nominated as: $\{G \to 0; 1/c \to 0; h \to 0\}$ and relativistic quantum mechanics as: $\{G \to 0; 1/c; h\}$. In this classification cosmological theories can be nominated as follows:

- Newtonian gravitational cosmology: $\{G; 1/c \to 0; h \to 0\}$
- Cosmology of the general relativity theory: $\{G; 1/c; h \to 0\}$
- Quantum cosmology: $\{G; 1/c; h\}$ with the equation of the universe evolution (9) represented, using Eq. 29, as a function of the fundamental constants of quantum physics (27).

New physical theories extending previous theoretical models, in accordance with Bohr "Correspondence Principle", usually can be reduced to previous models by some limit transformation. The concept of quantum cosmology is consistent with cosmological interpretation of Einstein equations of gravitation. The quantum equation (27) can be regarded as a special condition defining energy-momentum tensor of the cosmic fluid at the right part of Einstein equation of gravitation. To see it, one should first use Eq. 10 in the form: $\dot{a} = aa'$ to transform equation $\left(\dot{a}/a\right)^2 = 8\pi G/3c^2 \rho_E$ of the Standard cosmological model to physical time derivatives:

$$a'^2 = \frac{8\pi G}{3c^2} \rho_E \qquad (68)$$

Substituting in this equation one of the possible solutions of quantum cosmological model equation (11, 12): $a = (2H\tau)^{1/2}$, we obtain the relation:

$$\rho_E = \frac{3c^2 H}{16\pi G}\tau^{-1} \tag{69}$$

This relation differs from (27) only by the constant multiplier, demonstrating that the Standard cosmological model can be regarded as a specific form of the quantum model with constant average energy density instead of the non-stationary equation (27). In Standard cosmological model energy density has no explicit dependence on time and varies only due to increasing space volume. The quantum equation (27) can be interpreted as a description of the cosmological "long tail relaxation", typical of complex structures with memory, governing the universe evolution.

## References

1. Baryshev Yu., Teerikorpi P. Discovery of Cosmic Fractals. World Scientific Publishing Co. 2002.
2. Djorgovski S., Spinard H. Toward the application of a metric size function in galactic evolution and cosmology // Astrophys. J., 251, 417 - 423, 1981 (December 15).
3. Goldhaber G., Groom D.E., Kim A. et al. Timescale Stretch Parameterization of Type Ia Supernova B-band Light Curves // Astrophys. J. (in press); arXiv:astro-ph/0104382 v1 24 Apr 2001.
4. Heisenberg W. Physics and Philosophy. London: Allen & Unwin, 1963.
5. Hildebrandt S.R., Rebolo R., Rubino-Martin J.A. et al. COSMOSOMAS Observations of the CMB and Galactic Foregrounds at 11 GHz: Evidence for anomalous microwave emission at high Galactic Latitude // Mon. Not. R. Astron. Soc. (in press); arXiv: astro-ph / 0706 1873 13 Jun. 2007.
6. Kamke E. Differentialgleichungen. Losungsmethoden und Losungen. V.I. Gewohnliche Differentialgleichungen. Leipzig, 1959.
7. Labini S., Montuori M., Pietronero L. Scale-invariance of galaxy clustering // arXiv:astro-ph/9711073 v1 7 Nov 1997.
8. Landau L.D., Lifshitz E.M. Statistical Physics. 3rd Ed. Part 1. Oxford: Butterworth-Heinemann, 1996.
9. Peebles P.J.E. Principles of Physical Cosmology. Princeton NJ: Princeton University Press, 1993.
10. Sandage A. et al. The Hubble Constant: A Summary of the Hubble Space Telescope Program for the Luminosity Calibration of Type Ia Supernovae by Means of Cepheids // Astrophys. J. 653, 843 - 860 (2006).
11. SDSS DR6. http://www.sdss.org/dr6/ ; http://arxiv.org/abs/0707.3413 ; http://adsabs.harvard.edu/abs/2007arXiv0707.3413A
12. Synge J.L. Relativity: The General Theory. Amsterdam: 1960.
13. Taganov I.N. Discovery of Cosmological Deceleration of the Course of Time. St.-Petersburg: TIN. 2005. ISBN 5-902632-02-1.
14. Taganov I.N. Quantum Cosmology. Deceleration of Time. St.-Petersburg: TIN. 2008. ISBN 978-5-902632-05-4. На русском языке: Таганов И.Н. Квантовая космология. Замедление времени. СПб.: ТИН, 2008. ISBN 978-5-902632-04-8.
15. Taganov I.N. Thermal asymmetry of the cosmic microwave background radiation. Preprint. St.-Petersburg.: TIN, 2008. ISBN 978-5-902632-07-8
16. Козырев Н.А. Причинная или несимметричная механика в линейном приближении. Пулково, 1958.
17. Сорохтин О.Г., Ушаков С.А. Глобальная эволюция Земли. М.: Изд. МГУ, 1991.